\documentclass[10pt]{IEEEtran}
\usepackage{}
\usepackage{amssymb}
\usepackage{amsfonts}
\usepackage{epstopdf}
\usepackage{graphicx}
\usepackage{stmaryrd}
\usepackage{amssymb,amsmath}
\usepackage{multirow,url}
\usepackage{color,cite}
\usepackage{hyperref}

\ifodd 0
\newcommand{\com}[1]{\textbf{\color{red} (COMMENT: #1)}} %comment of the text
\newcommand{\comg}[1]{\textbf{\color{green} (COMMENT: #1)}}
\newcommand{\response}[1]{\textbf{\color{magenta} (RESPONSE: #1)}} %response to comment
\else

\newcommand{\com}[1]{}
\newcommand{\comg}[1]{}
\newcommand{\response}[1]{}
\fi
\newtheorem{definition}{Def\/inition}

\ifCLASSINFOpdf
\else
\fi
\begin{document}
%
% paper title
% can use linebreaks \\ within to get better formatting as desired
%\title{Asynchronous Physical-layer Network Coding}
\title{Optimal Decoding Algorithm for Asynchronous Physical-Layer Network Coding}
% author names and affiliations
% use a multiple column layout for up to three different
% affiliations
%\author{\IEEEauthorblockN{Lu Lu}
%\IEEEauthorblockA{Dept. of Information Engineering\\The Chinese University of Hong Kong\\
%Email: ll007@ie.cuhk.edu.hk} \and \IEEEauthorblockN{Soung Chang
%Liew}
%\IEEEauthorblockA{Dept. of Information Engineering\\The Chinese University of Hong Kong\\
%Email: soung@ie.cuhk.edu.hk}}

%\author{Lingjie Duan$^{\ast}$, Jianwei Huang$^{\ast}$, Biying Shou$^{\dagger}$\\$^{\ast}$
%Department of Information Engineering, The Chinese University of Hong Kong, Hong Kong\\
%$^{\dagger}$Department of Management Sciences, City University of Hong Kong, Hong Kong\\
%email:\{dlj008,jwhuang\}@ie.cuhk.edu.hk,biying.shou@cityu.edu.hk
%
%\thanks{This work is supported by the Competitive Earmarked
%Research.}}

%\author{Lu Lu and Soung Chang Liew\\
%Department of Information Engineering, The Chinese University of Hong Kong, Hong Kong\\
%email:\{ll007, soung\}@ie.cuhk.edu.hk
%
%%\thanks{This work was supported by the Competitive Earmarked Research Grant (Project ID 414507)
%%established under the University Grant Committee of the Hong Kong
%%Special Administrative Region, China, and the Direct Grant (Project
%%ID 2050436) of the Chinese University of Hong Kong. This work is
%%also supported by National Science Foundation of China (Grant No.
%%60902016).}
%}

\author{Lu Lu$^{\ast}$, Soung Chang Liew$^{\ast}$ and Shengli Zhang$^{\ast}$$^{\dagger}$\\
$^{\ast}$Department of Information Engineering, The Chinese University of Hong Kong, Hong Kong\\
$^{\dagger}$Department of Communication Engineering, Shenzhen University, China\\
email:\{ll007, soung\}@ie.cuhk.edu.hk, zsl@szu.edu.cn
\thanks{This work was partially supported by grants from the University Grants Committee of the Hong Kong Special Administrative Region, China (Project No. AoE/E-02/08; Project No. 2150537). This work was also supported by NSF of China (Project No. 60902016) and NSF of Guangdong (Project No. 10151806001000003).}}

% conference papers do not typically use \thanks and this command
% is locked out in conference mode. If really needed, such as for
% the acknowledgment of grants, issue a \IEEEoverridecommandlockouts
% after \documentclass
% use for special paper notices
%\IEEEspecialpapernotice{(Invited Paper)}
% make the title area

\maketitle

\begin{abstract}
%\boldmath

A key issue in physical-layer network coding (PNC) is how to deal with the asynchrony between signals transmitted by multiple transmitters. That is, symbols transmitted by different transmitters could arrive at the receiver with symbol misalignment as well as relative carrier-phase offset. In this paper, 1) we propose and investigate a general framework based on belief propagation (BP) that can effectively deal with symbol and phase asynchronies; 2) we show that for BPSK and QPSK modulations, our BP method can significantly reduce the SNR penalty due to asynchrony compared with prior methods; 3) we find that symbol misalignment makes the system performance less sensitive and more robust against carrier-phase offset. Observation 3) has the following practical implication. It is relatively easier to control symbol timing than carrier-phase offset. Our results indicate that if we could control the symbol offset in PNC, it would actually be advantageous to deliberately introduce symbol misalignment to desensitize the system to phase offset.

%A key issue in physical-layer network coding (PNC) is how to deal with the asynchrony between signals transmitted by multiple transmitters. In wireless networks, symbols transmitted by different transmitters could arrive at the receiver with symbol misalignment as well as relative carrier-phase offset even with synchronization mechanism. In this paper, 1) we propose and investigate a general framework based on belief propagation (BP) that can effectively deal with symbol and phase asynchronies; 2) we show that for BPSK and QPSK modulations, our BP method can significantly reduce the SNR penalty due to asynchrony compared with prior methods; 3) we find that symbol misalignment makes the system performance less sensitive and more robust against carrier-phase offset. Observation 3) has the following practical implication. It is relatively easier to control symbol timing than carrier-phase offset. Our results indicate that if we could control the symbol offset in PNC, it would actually be advantageous to deliberately introduce symbol misalignment to desensitize the system to phase offset.

\end{abstract}
% IEEEtran.cls defaults to using nonbold math in the Abstract.
% This preserves the distinction between vectors and scalars. However,
% if the conference you are submitting to favors bold math in the abstract,
% then you can use LaTeX's standard command \boldmath at the very start
% of the abstract to achieve this. Many IEEE journals/conferences frown on
% math in the abstract anyway.

% no keywords

% For peer review papers, you can put extra information on the cover
% page as needed:
% \ifCLASSOPTIONpeerreview
% \begin{center} \bfseries EDICS Category: 3-BBND \end{center}
% \fi
%
% For peerreview papers, this IEEEtran command inserts a page break and
% creates the second title. It will be ignored for other modes.
\IEEEpeerreviewmaketitle

\section{Introduction}
% no \IEEEPARstart
%This demo file is intended to serve as a ``starter file''
%for IEEE conference papers produced under \LaTeX\ using
%IEEEtran.cls version 1.7 and later.
%% You must have at least 2 lines in the paragraph with the drop letter
%% (should never be an issue)
%I wish you the best of success.
%
%\hfill mds
%
%\hfill January 11, 2007

Physical-layer network coding (PNC), first proposed in \cite{PNC06}, is a subfield of network coding \cite{AhlswIT00} that is attracting much attention recently. The simplest system in which PNC can be applied is the two-way relay channel (TWRC), where two end nodes exchange information with the help of a relay node in the middle, as illustrated in Fig. \ref{fig:system}(a).

Compared with the conventional relay system, PNC doubles the throughput of TWRC by reducing the needed time slots from four to two. In PNC, in the first time slot, the two end nodes send signals simultaneously to the relay; in the second phase, the relay processes the superimposed signals and maps them to network-coded information for broadcast back to the end nodes. From the network-coded information, each end node then makes use of its self information to extract the information from the other end node.
%\footnote{Note that PNC is different from Multi-user Detection (MUD) \cite{Verdubook}. In PNC, the relay attempts to decode the \emph{XOR} of the symbols from two users, whereas in MUD the individual symbols from the users are to be decoded.}

A key issue in PNC is how to deal with the asynchrony between signals transmitted by multiple transmitters. That is, symbols transmitted by different transmitters could arrive at the receiver with symbol misalignment as well as relative carrier-phase offset.

Many previous works (\emph{e.g.}, \cite{PNC06, SyncPNC06, HaoMilcom07}) found that symbol misalignment and carrier-phase offset will result in appreciable performance penalties. For BPSK modulation, \cite{PNC06} showed that the BER performance penalties due to the carrier phase asynchrony and symbol offset are both 3 dB in the worst case. For QPSK modulation, the penalty can be as large as 6 dB in the worst case when the carrier-phase offset is $\pi/4$ \cite{HaoMilcom07}.

These earlier investigations led to a common understanding that near-perfect symbol and carrier-phase synchronization are important for good performance in PNC. We believe that this is a misconception, and that asynchronous PNC can have good performance if we use the right methods to deal with the asynchrony.

The study of BPSK PNC in \cite{PNC06, SyncPNC06}, for example, made use of suboptimal decoding methods at the relay for the asynchronous PNC case. Furthermore, the joint effect of symbol and phase asynchronies was not investigated. In this paper, we propose an optimal maximum-likelihood (ML) decoding method making use of a belief propagation (BP) algorithm that addresses symbol and phase asynchronies jointly. We find that the BER performance penalty can be reduced from 3 dB (using the method in \cite{PNC06, SyncPNC06}) to less than 0.5 dB (using our method here) for the worst case.

In QPSK PNC, the penalty is larger than 6 dB \cite{HaoMilcom07} only when the symbols are perfectly aligned and when the phase offset is $\pi/4$. We find that using our optimal BP decoding algorithm, when there is a 1/2 symbol misalignment, the penalty is less than 1 dB (compared with the case when symbol and phase are perfectly synchronized). Additionally, an interesting result is that with the 1/2 symbol misalignment, the spread of penalties under various carrier-phase offsets is no more than 0.5 dB. This means that symbol misalignment has the effect of desensitizing the performance of the system to carrier-phase offset. The practical implication is as follows:

%We will show that using our optimal BP decoding algorithm, the penalty is much reduced when the symbols are misaligned. In particular, when
%there is a 1/2 symbol misalignment, the spread of penalties dues to different phase offsets is only less than 0.5 dB \com{insert
%spread of penalty here in terms of dB}; furthermore, compared with the perfectly synchronized PNC with zero symbol and zero phase
%offsets, the largest penalty is no more than 1dB \com{insert result here}.

In practice, it is easier to control symbol timing than carrier-phase offset because of the relative time scales involved. If we could control the symbol offset in PNC (\emph{e.g.}, \cite{RahulSourceSync10} presented a method to synchronize symbols from different sources), it would actually be advantageous to deliberately introduce symbol misalignment so that the system is robust against uncontrollable phase offsets.

The remainder of this paper is organized as follows: Section II introduces the system model adopted by this paper. Section III classifies and defines synchronous PNC and asynchronous PNC. Section IV presents our BP ML decoding method for dealing with asynchrony in PNC. Numerical results are given in Section V. Finally, Section VI concludes this paper.

\section{System Model}
%\vspace*{-.094in}

\begin{figure}[tt]
\centering
\includegraphics[width=0.4\textwidth]{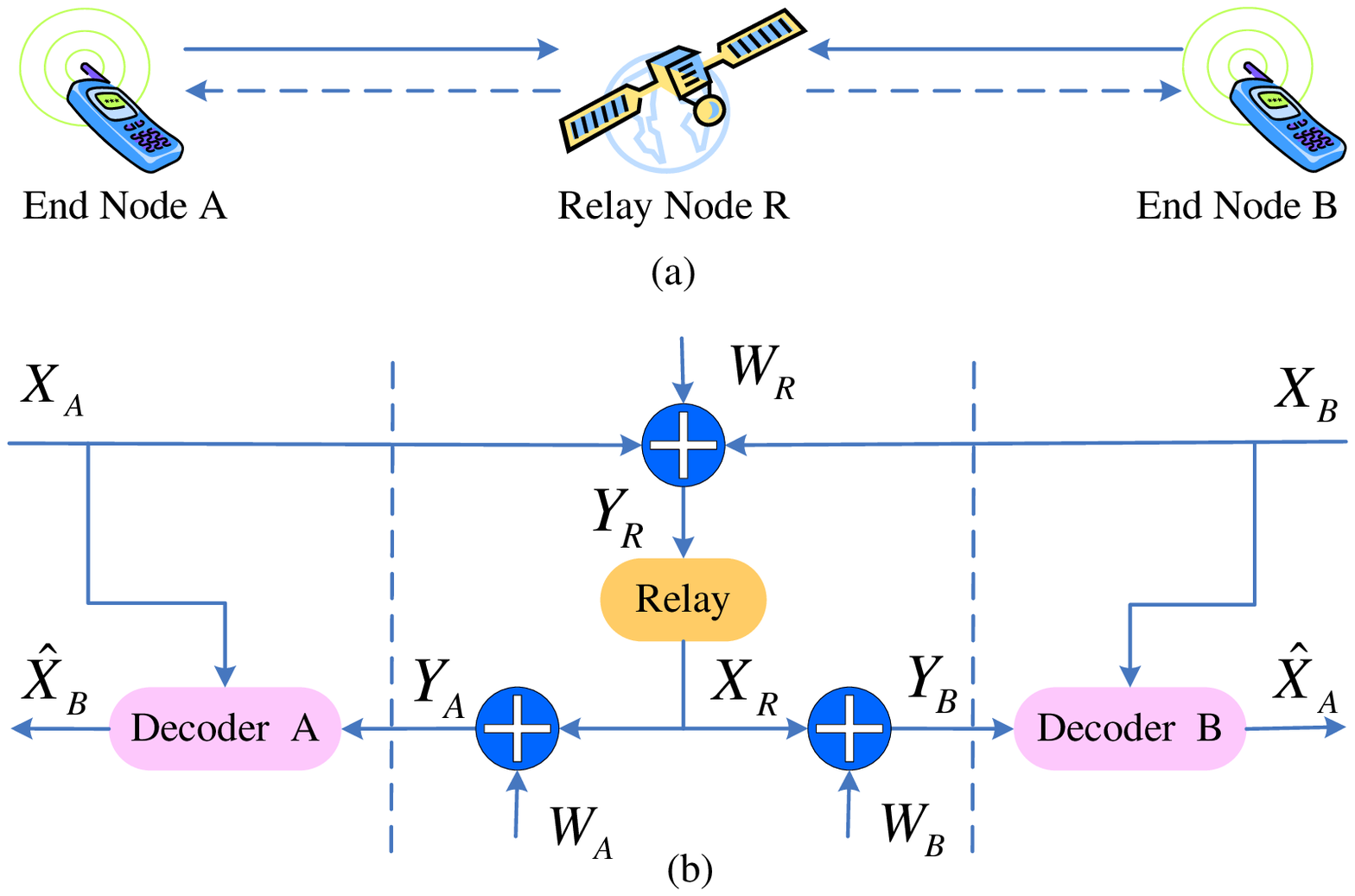}
\caption{System model for two way relay channel.} \label{fig:system}
\end{figure}

We study the two-way relay channel as shown in Fig. \ref{fig:system}, in which nodes \emph{A} and \emph{B} exchange information with the help of relay node \emph{R}. We assume that all nodes are half-duplex, \emph{i.e.}, a node cannot receive and transmit simultaneously. We also assume that there is no direct link between nodes \emph{A} and \emph{B}.
%An example in practice is a satellite communication system in which the two end nodes on the earth can only communicate with each other via the relay satellite, as shown in Fig. \ref{fig:system}(a).

We consider a two-phase transmission scheme consisting of an uplink phase and a downlink phase. In the uplink phase, nodes \emph{A} and
\emph{B} transmit packets to node \emph{R} simultaneously. In the downlink phase, based on the overlapped signals received from \emph{A} and \emph{B}, \emph{R} constructs a network-coded packet and broadcast the packet to \emph{A} and \emph{B}. Upon receiving the network-coded packet, \emph{A} (\emph{B}) then attempts to recover the original packet transmitted by \emph{B} (\emph{A}) in the uplink phase using self-information \cite{PNC06}.

Consider the uplink phase again, if only \emph{A} transmitted, then the received complex baseband signal at \emph{R} (\emph{i.e.}, the received signal after down-conversion from the carrier frequency and low-pass filtering) would be
\begin{align}
y_R (t) = \sum\limits_{n = 1}^N {h_A x_A [n]p(t - nT)}  + w_R (t),
\label{equ:ApncSys1}
\end{align}
where $h_A  = \sqrt {P_A } $ is the received signal amplitude; $(x_A [n])_{n = 1,...,N} $ are the symbols in the packet of \emph{A}; $p(t
- nT)$ is the pulse shaping function for the baseband signal; and $w_R (t)$ is Gaussian noise with double-sided power spectral density $S_w (f) =
{{N_0 } \mathord{\left/{\vphantom {{N_0 } 2}} \right. \kern-\nulldelimiterspace} 2}$ for both the real and imaginary components within the baseband of interest. In the PNC set-up, \emph{A} and \emph{B} transmit simultaneously. In this case, the received complex baseband signal at \emph{R} is
\begin{align}
y_R (t) = &\sum\limits_{n = 1}^N {\left\{ {h_A x_A [n]p(t - nT) + h_B
x_B [n]p(t - \Delta  - nT)} \right\}}\nonumber\\  &+ w_R (t),
\label{equ:ApncSys2}
\end{align}
where $h_B  = \sqrt {P_B } e^{j\phi }$ ($\phi$ is the relative phase offset between the signals from \emph{A} and \emph{B} due to phase asynchrony in their carrier-frequency oscillators, and difference in the delays of the two uplink channels); $(x_B [n])_{n = 1,...,N}$ are
the symbols in the packet of \emph{B}; and $\Delta$ is a time offset between the arrivals of the signals from \emph{A} and \emph{B}. Without loss of generality, we assume the signal of \emph{A} arrives earlier than \emph{B}. Furthermore, we assume $\Delta$ is within one symbol period \emph{T}. Thus, $0 \le \Delta  < T$.\footnote{If $\Delta$ is more than one symbol period, we could generalize our treatment here so that $N$ is larger than the number of symbols in a packet. The packets will only be partially overlapping, with non-overlapping symbols at the front end and tail end. Essentially, our assumption of $\Delta$ being within one symbol period implies that we are looking at the ``worst case'' with maximum overlapping between the two packets. When there are additional non-overlapping symbols at the front and tail ends, the decoding will have better error probability performance.} We refer to $\Delta$ and $\phi$ as the symbol and phase offsets (or misalignments) at \emph{R}, respectively. Note that we have assumed a flat fading channel, and the channel gains stay constant within a packet duration. For simplicity, we assume power control so that $P_A  = P_B  = P$. Furthermore, for convenience, we assume time is expressed in unit of symbol duration, so that $T = 1$. Then, we can rewrite (\ref{equ:ApncSys2}) as
\begin{align}
y_R (t) = &\sqrt P \sum\limits_{n = 1}^N {\left\{ {x_A [n]p(t - n) + x_B [n]p(t - \Delta  - n)e^{j\phi } } \right\}}\nonumber\\  &+ w_R (t).
\label{equ:ApncSys3}
\end{align}

In general, the pulse shaping function $p(t)$ can take different forms. To bring out the essence of our results in the simplest manner, throughout this paper, we assume the rectangular pulse shape: $p(t) = rect(t) = u(t + 1) - u(t)$.

A critical design issue is how node \emph{R} makes use of $y_R (t)$ to construct a packet for broadcast to nodes \emph{A} and \emph{B}
in the downlink phase. In this paper, we assume that \emph{R} first oversamples $y_R (t)/\sqrt P$ to obtain $2N+1$ signal samples. The $2N+1$ signal samples will be used to construct an \emph{N}-symbol packet for broadcast to \emph{A} and \emph{B}. The oversampling procedure is similar to that in \cite{CCRESM} and described below.
For $n=1,...,N,$
\begin{align}
y_R [2n - 1] &= \frac{1}{\Delta}\int_{(n - 1)}^{(n - 1) + \Delta } {( {x_A [n] + x_B [n - 1]e^{j\phi }  + \frac{w_R (t)}{\sqrt{P}}} )d} t \nonumber\\
&= x_A [n] + x_B [n - 1]e^{j\phi }  + w_R [2n - 1], \nonumber\\
y_R [2n] &= \frac{1}{{1 - \Delta } }\int_{(n - 1) + \Delta }^n {( {x_A [n] + x_B [n]e^{j\phi }  + \frac{w_R (t)}{\sqrt{P}}} )d} t
\nonumber\\
&= x_A [n] + x_B [n]e^{j\phi }  + w_R [2n], \label{equ:ApncSys4} \\
\text{and}\ \ \ \ \ \ \ \ \ \ &\ \nonumber\\
y_R [2N + 1] &= \frac{1}{\Delta \sqrt{P}}\int_N^{N + \Delta } {\left( {x_B [N]e^{j\phi }  + w_R (t)/\sqrt{P}} \right)d} t \nonumber\\
&= x_B [N]e^{j\phi }  + w_R [2N + 1], \nonumber
\end{align}
where $x_B [0] = 0$, and $w_R [2n-1]$ (also $w_R [2N+1]$) and $w_R [2n]$ are a zero-mean complex Gaussian noise with variance $N_0 /(2P\Delta )$ and $N_0 /\left( {2P(1 - \Delta )} \right)$, respectively, for both the real and imaginary components.

Fig. \ref{fig:system}(b) is a schematic diagram of our system. This paper adopts the following notation:
\begin{itemize}
%\item $S_i  = \left( {s_i [1],s_i [2],...,s_i [M]} \right)$ denotes the source packet of node \emph{i}, $i \in \{ A,B\}$;

\item $X_i  = \left( {x_i [1],x_i [2],...,x_i [N]} \right)$ denotes the source packet of node \emph{i}, $i \in \{ A,B\}$;

\item $Y_R  = \left( {y_R [1],y_R [2],...,y_R [N],y_R [N + 1],...,y_R [2N + 1]} \right)$ denotes the received packet (with the aforementioned oversampling) at relay node \emph{R};

\item $W_R  = \left( {w_R [1],w_R [2],...,w_R [N],w_R [N + 1],...,\\w_R [2N + 1]} \right)$ denotes the receiver noise at node \emph{R};

\item $X_R  = \left( {x_R [1],x_R [2],...,x_R [N]} \right)$ denotes the network-coded packet at relay node \emph{R};

\item $Y_i  = \left( {y_i [1],y_i [2],...,y_i [N]} \right)$ denotes the received PNC packet at node \emph{i}, $i \in \{ A,B\}$;

\item $W_i  = \left( {w_i [1],w_i [2],...,w_i [N]} \right)$ denotes the receiver noise at node \emph{i}, $i \in \{ A,B\}$;

\item $\hat X_i  = \left( {\hat x_i [1],\hat x_i [2],...,\hat x_i [N]} \right)$ denotes the decoded source packet of node \emph{i}, $i \in \{A,B\}$, at the other end node;
\end{itemize}
where $N$ is the number of source symbols. In the above, for BPSK, $x_i [ \cdot ],\hat x_i [ \cdot ],x_R [ \cdot ] \in \{-1,1\}$; for QPSK, $x_i [ \cdot ],\hat x_i [ \cdot ],x_R [ \cdot ] \in \{ {{(1 + j)} \mathord{\left/ {\vphantom {{(1 + j)} {\sqrt 2 }}} \right.
 \kern-\nulldelimiterspace} {\sqrt 2 }},{{{\rm{ }}( - 1 + j)} \mathord{\left/ {\vphantom {{{\rm{ }}( - 1 + j)} {\sqrt 2 ,}}} \right.
 \kern-\nulldelimiterspace} {\sqrt 2 ,}}{\rm{ }}{{( - 1 - j)} \mathord{\left/ {\vphantom {{( - 1 - j)} {\sqrt 2 ,}}} \right.
 \kern-\nulldelimiterspace} {\sqrt 2 ,}}{\rm{ }}{{(1 - j)} \mathord{\left/ {\vphantom {{(1 - j)} {\sqrt 2 }}} \right. \kern-\nulldelimiterspace} {\sqrt 2 }}\}$
. In addition, $y_i [ \cdot ],y_R [ \cdot ],w_i [ \cdot ],w_R [ \cdot ] \in \mathbb{C}$.

\section{Synchronous vs. Asynchronous PNC}
\subsection{Classification} \label{Classification}
Table \ref{tab:Class} shows the four possible cases in PNC systems. Case 1 is the perfectly synchronized case studied in \cite{PNC06}; Case 2, in \cite{SyncPNC06}; and Case 3, in \cite{SyncPNC06, ShengliGlobcom08, KoikeJSAC09}. This paper proposes a general scheme to tackle all four cases under one framework.
\vspace*{.08in}
\begin{table}[h]
\caption{Four cases of PNC systems}
\centering
\begin{tabular}{|c|c|c|}
\hline \multicolumn{1}{|c|}{} &\multicolumn{1}{|c|}{$\Delta  = 0$} &\multicolumn{1}{|c|}{$\Delta  \ne 0$}
\\\hline $\phi  = 0$ & $\phi  = 0,\Delta  = 0$ (Case 1) & $\phi  = 0,\Delta  \ne 0$ (Case 2)
\\\hline $\phi  \ne 0$ & $\phi  \ne 0,\Delta  = 0$ (Case 3) & $\phi \ne 0,\Delta  \ne 0$ (Case 4)
\\\hline
\end{tabular}
\tabcolsep -10mm\label{tab:Class}
\end{table}

%This section is organized as follows: First we give the definition and a brief review of synchronous PNC. Then we extend to asynchronous PNC. Finally we present, BP-PNC, targeting for solving general asynchrony problems (\emph{i.e.}, symbol and phase offsets) in PNC.

%In this section, we construct C-CRESM based on RA (repeat
%accumulate) code. This section is organized as follows: First we
%give a brief review of the RA code with its decoding algorithm.
%Interested readers who wish to learn more about RA code are referred
%to \cite{RAconf} for details. Then we construct a \emph{virtual}
%Tanner graph, that reveals our C-CRESM can integrate channel coding
%with collision resolution. Finally, the message update rules of the
%associate belief propagation algorithm are derived to compute the
%source information of both nodes, based on the \emph{virtual} Tanner
%graph.

\subsection{Review of Synchronous PNC (Case 1 in \ref{Classification})}
\begin{definition}[Synchronous PNC]
Synchronous PNC is the case where the two end nodes transmit their packets $X_A$ and $X_B$ in a synchronous manner so that the relay node \emph{R} receives the combined signals with $\phi  = 0$ and $\Delta  = 0$. The received baseband packet at \emph{R} is $Y_R  = X_A  + X_B  + W_R$ with \emph{N} symbols. Node \emph{R} then transforms $Y_R$ into a network-coded packet $X_R  = f(Y_R )$ for transmission in the downlink phase.
\end{definition}

For this case, with reference to (\ref{equ:ApncSys4}), since $\Delta  = 0$, the variance of the noise term $w_R [2n - 1]$ is infinite, and the signal is contained only in the even terms $y_R [2n]$. Thus, we can write
\begin{align}
y_R [2n] = x_A [n] + x_B [n] + w_R [2n], \label{equ:ApncBP1}
\end{align}
where $n = 1,...,N$, and $w_R [2n]$ is zero-mean Gaussian noise with variance $\sigma ^2  = N_0 /2$ for both the real and imaginary components.

For BPSK, $x_i [n] \in \{- 1,1\}$. Only the real component of $w_R [2n]$ needs to be considered. For QPSK, since we are considering a synchronous system, the in-phase and quadrature-phase components in (\ref{equ:ApncBP1}) are independent; it can therefore be considered as two parallel BPSK systems. In the following, we assume BPSK.

Let us consider a particular \emph{n}, and omit the index \emph{n} in our notation for simplicity. The \emph{a posteriori} probability of the combination $(x_A ,x_B )$ is given by
\begin{align}
&\Pr (x_A ,x_B |y_R ) = \frac{{\Pr (y_R |x_A ,x_B )}}{{4\Pr (y_R )}} \nonumber\\
&= \frac{1}{{4\Pr (y_R )\sqrt {2\pi \sigma ^2 } }}\exp \left\{ {\frac{{[y_R  - x_A  - x_B )]^2 }}{{2\sigma ^2 }}} \right\}.
\end{align}

Let us use $x_i  = 1$ to represent bit 0 and $x_i  = -1$ to represent bit 1. Suppose that the downlink transmission also uses BPSK. For PNC output, $x_R $, let us consider the \emph{XOR} mapping: $x_R  = x_A  \oplus x_B $ \cite{PNC06}. Then, $x_R  = 1$ if $x_A  = x_B $, and $x_R  =  - 1$ if $x_A  \ne x_B $. The following decision rule is used to map $y_R $ to $x_R $:
\begin{align}
&\Pr (x_A  = 1,x_B  = 1|y_R ) + \Pr (x_A  =  - 1,x_B  =  - 1|y_R )
\mathop \gtrless \limits^{x_R  = 1} \limits_{x_R  = -1} \nonumber\\ &\Pr (x_A  = 1,x_B  =  - 1|y_R ) + \Pr (x_A  =  - 1,x_B  = 1|y_R ) \nonumber\\
\Rightarrow
&\left( {\exp \left\{ {\frac{{(y_R  - 2)^2 }}{{2\sigma ^2 }}} \right\} + \exp \left\{ {\frac{{(y_R  + 2)^2 }}{{2\sigma ^2 }}} \right\}} \right) \mathop \gtrless \limits^{x_R  = 1} \limits_{x_R  = -1} \exp \left\{ {\frac{{y_R ^2 }}{{2\sigma ^2 }}} \right\}. \label{equ:ApncBP2}
\end{align}

%%%%%%%%%%%%%%%%%%%%%%%%

\subsection{Asynchronous PNC (Case 2, 3 and 4 in \ref{Classification})}
\begin{definition}[Asynchronous PNC]
Asynchronous PNC is the case where the two end nodes transmit their packets $X_A$ and $X_B$ in an asynchronous manner so that the relay node \emph{R} receives the combined signals with $\phi \ne 0$ or $\Delta \ne 0$. The received baseband packet at \emph{R} is $Y_R  = X_A  + X_B  + W_R$ with $2N+1$ symbols. Node \emph{R} then transforms $Y_R$ into a network-coded packet $X_R  = f(Y_R )$ for transmission in the downlink phase.
\end{definition}

The asynchronous case has been studied previously. However, suboptimal PNC mappings $X_R  = f(Y_R )$ were considered. Refs. \cite{PNC06} and \cite{SyncPNC06} argued that the largest BER performance penalty is 3 dB for BPSK modulation (for both phase and symbol asynchronies). These results, however, are based on suboptimal decoding methods. Ref. \cite{HaoMilcom07} mentioned that there is a maximum 6 dB BER performance penalty for QPSK modulation when $\Delta = 0$ and $\phi \ne 0$. To the best of our knowledge, no quantitative results and concrete explanation have been given for the QPSK case for general $\Delta$ and $\phi$.

Ref. \cite{KattiANC07} used an amplify-and-forward method in which the overlapped signals received at the relay are simply amplified and broadcasted back to the end nodes. This method does not require synchronization between the end nodes, but introduces some inefficiency, because the relay's noise is forwarded along with the signal to the end nodes. Ref. \cite{KoikeJSAC09} investigated systems in which symbols are aligned but phases are not. It uses QPSK for uplinks, but a higher order constellation map (\emph{e.g.}, 5QAM) for downlinks when the uplink phase offset is such that the use of QPSK will lead to poor performance. Ref. \cite{ZorziSPAWC09} investigated OFDM PNC. Because of the long duration of OFDM symbols in the frequency domain, the symbols from two end nodes can be considered as ``almost aligned''. However, the phase offset is not addressed.

%The phase issue is dealt with via diversity across the different OFDM subcarriers, in which some subcarriers have phase offset that leads to good performance, and some have phase offset that leads to poor performance.

None of the prior methods attempt to look at the symbol and phase offset jointly. Furthermore, they try to ``deal with'' asynchrony, rather than to ``exploit'' asynchrony to gain performance advantage. As will be shown in this paper, we can exploit symbol asynchrony to gain robustness of the system against phase asynchrony.

%However, all of the above methods inevitably introduces some kind of redundancy, and also are methods that try to avoid asynchrony rather than exploiting it. In the next section we will introduce a method that focuses on the case where both the uplink and downlink phases use QPSK, to solve the asynchrony problem in PNC.
%\textcolor{blue}{A reason is that we will consider channel-coded PNC systems in the future, in which if both the uplink and downlink phases use the same modulation, the channel coding and PNC mapping operation at the relay node will be simpler. (Due to limited space in this paper, we will not present the channel-coded PNC cases.)} For easy comparison and uniformity, we therefore make the same assumption here.

\section{A Belief Propagation PNC Decoding Scheme (BP-PNC)}
This section presents a PNC decoding scheme based on belief propagation that deals with symbol and phase asynchronies jointly. We refer to the decoding method as BP-PNC. We make use of the oversampled symbols in (\ref{equ:ApncSys4}) to construct a Tanner graph \cite{KschIT01} as shown in Fig. \ref{fig:TannerBPPNC}. In the Tanner graph, $Y_R$ denotes the evidence nodes, and there are $2N+1$ such nodes; $\Psi$ denotes the constraint nodes (also known as the compatibility or check nodes); and \emph{X} denotes the source (or variable) nodes. We aim to decode the combination $\left( {x_A [n],x_B [n]} \right)$ in $X$. For simplicity, we use $x^{i,j}$ to denote $\left( {x_A [i],x_B [j]} \right)$. Note that the Tanner graph has a tree structure. This means that BP can find the exact solution for the \emph{a posteriori} probability $P(x^{i,j} |Y_R )$ at each source node. Furthermore, the solution can be found after only one iteration of the message-passing algorithm (see \cite{KschIT01}). From the decoded $P(x^{n,n} |Y_R)$, we can then find the maximum \emph{a posteriori} probability (MAP) \emph{XOR} value, $x_R [n] = \arg \mathop {\max }\limits_{x_R [n]} \sum\limits_{x^{n,n} :\ x_A [n] \oplus x_B [n] = x_R [n]} {\Pr \left( {x^{n,n} |Y_R } \right)}$. In summary, BP can converge quickly and is MAP-optimal as far as the BER of $x_A [n] \oplus x_B [n]$ is concerned. Note that the MAP optimal scheme is also ML optimal because each of the possible symbol values for $x_A [n]$ and $x_B [n]$ are equally likely (\emph{i.e.}, $\Pr \left( {x_A [n] = \eta _A } \right) = \Pr \left( {x_B [n] = \eta _B } \right) = \frac{1}{{|\chi |}}$, where ${\eta_A}, {\eta_B} \in \chi$, and $\chi$ is the alphabet of modulated symbols.).

\subsubsection{BP-PNC Design}
Let us now consider QPSK modulation, in particular, we define $\chi = \left\{ 1 + j, - 1 + j, - 1 - j,1 - j \right\}$ as the symbol set. With reference to (\ref{equ:ApncSys4}), we have $x_A [n] = {a \mathord{\left/ {\vphantom {a {\sqrt 2 }}} \right. \kern-\nulldelimiterspace} {\sqrt 2 }}$ and $x_B [n] = {b \mathord{\left/ {\vphantom {b {\sqrt 2 }}} \right. \kern-\nulldelimiterspace} {\sqrt 2}}$, where $a,b \in \chi$. Define $p_k^{a,b}  = P\left( {x_A [\left\lceil {k/2} \right\rceil ] = {a \mathord{\left/
 {\vphantom {a {\sqrt 2 }}} \right.
 \kern-\nulldelimiterspace} {\sqrt 2 }},x_B [\left\lfloor {k/2} \right\rfloor ] = {b \mathord{\left/
 {\vphantom {b {\sqrt 2 }}} \right.
 \kern-\nulldelimiterspace} {\sqrt 2 }}|y_R [k]} \right)$. Note that here, $p_k^{a,b}$ is computed based on $y_R [k]$ only, and not on the whole $Y_R $. Also, $p_k^{a,b}$ is fixed and does not change throughout the message passing algorithm in the Tanner graph. $P_{2n - 1}^{a,b}$ and $P_{2n}^{a,b}$ for $n = 1,2,...,N$ are given as follows:
\begin{align}
&p_{2n - 1}^{a,b}  = P\left(x_A [n] = \frac{a}{{\sqrt 2 }},x_B [n - 1] = \frac{b}{{\sqrt 2 }} \bigg\vert y_R [2n - 1] \right) \nonumber\\
&= \frac{1}{{2\pi \sigma ^2 /\Delta }}\exp \left\{ {\frac{{\left( {y_R^{{\mathop{\rm Re}\nolimits} } [2n - 1] - {{{\mathop{\rm Re}\nolimits} \left( {a + be^{j\phi } } \right)} \mathord{\left/
{\vphantom {{{\mathop{\rm Re}\nolimits} \left( {a + be^{j\phi } } \right)} {\sqrt 2 }}} \right.
\kern-\nulldelimiterspace} {\sqrt 2 }}} \right)^2 }}{{2\sigma ^2 /\Delta }}} \right\}\cdot \nonumber\\
&\ \ \ \ \exp \left\{ {\frac{{\left( {y_R^{{\mathop{\rm Im}\nolimits} } [2n - 1] - {{{\mathop{\rm Im}\nolimits} \left( {a + be^{j\phi } } \right)} \mathord{\left/
{\vphantom {{{\mathop{\rm Im}\nolimits} \left( {a + be^{j\phi } } \right)} {\sqrt 2 }}} \right.
\kern-\nulldelimiterspace} {\sqrt 2 }}} \right)^2 }}{{2\sigma ^2 /\Delta }}} \right\}, \label{equ:ApncBPIni1}
\end{align}
\begin{align}
&p_{2n}^{a,b}  = P\left(x_A [n] = \frac{a}{{\sqrt 2 }},x_B [n] = \frac{b}{{\sqrt 2 }} \bigg\vert y_R [2n] \right) \nonumber\\
&= \frac{1}{{2\pi \sigma ^2 /(1 - \Delta )}}
\exp \left\{ {\frac{{\left( {y_R^{{\mathop{\rm Re}\nolimits} } [2n] - {{{\mathop{\rm Re}\nolimits} \left( {a + be^{j\phi } } \right)} \mathord{\left/
{\vphantom {{{\mathop{\rm Re}\nolimits} \left( {a + be^{j\phi } } \right)} {\sqrt 2 }}} \right.
\kern-\nulldelimiterspace} {\sqrt 2 }}} \right)^2 }}{{2\sigma ^2 /(1 - \Delta )}}} \right\}\cdot \nonumber\\
&\ \ \ \ \exp \left\{ {\frac{{\left( {y_R^{{\mathop{\rm Im}\nolimits} } [2n] - {{{\mathop{\rm Im}\nolimits} \left( {a + be^{j\phi } } \right)} \mathord{\left/
{\vphantom {{{\mathop{\rm Im}\nolimits} \left( {a + be^{j\phi } } \right)} {\sqrt 2 }}} \right.
\kern-\nulldelimiterspace} {\sqrt 2 }}} \right)^2 }}{{2\sigma ^2 /(1 - \Delta )}}} \right\}. \label{equ:ApncBPIni2}
\end{align}
Note that except for the first and last symbols, each of $p_{2n - 1}^{a,b}$ and $p_{2n}^{a,b}$ has 16 possible combinations (4 possibilities for $a$ and 4 possibilities for $b$). The first and last symbols have 4 possibilities, as follows:
$ p_1^{a,0}  = P(x_A [N + 1] = {a \mathord{\left/
{\vphantom {a {\sqrt 2 }}} \right.
\kern-\nulldelimiterspace} {\sqrt 2 }},x_B [n] = 0|y_R [1])
= \frac{1}{{2\pi \sigma ^2 /\Delta }}\exp \left\{ {\frac{{\left( {y_R^{{\mathop{\rm Re}\nolimits} } [1] - {\mathop{\rm Re}\nolimits} \left( {{a \mathord{\left/
{\vphantom {a {\sqrt 2 }}} \right.
\kern-\nulldelimiterspace} {\sqrt 2 }}} \right)} \right)^2 }}{{2\sigma ^2 /\Delta }}} \right\}
\exp \left\{ {\frac{{\left( {y_R^{{\mathop{\rm Im}\nolimits} } [1] - {\mathop{\rm Im}\nolimits} \left( {{a \mathord{\left/
{\vphantom {a {\sqrt 2 }}} \right.
\kern-\nulldelimiterspace} {\sqrt 2 }}} \right)} \right)^2 }}{{2\sigma ^2 /\Delta }}} \right\},$
and $ p_{2N + 1}^{0,b}  = P(x_A [N + 1] = 0,x_B [n] = {b \mathord{\left/
{\vphantom {b {\sqrt 2 }}} \right.
\kern-\nulldelimiterspace} {\sqrt 2 }}|y_R [2N + 1])
= \frac{1}{{2\pi \sigma ^2 /\Delta }} \exp \left\{ {\frac{{\left( {y_R^{{\mathop{\rm Re}\nolimits} } [2N + 1] - {\mathop{\rm Re}\nolimits} \left( {{{be^{j\phi } } \mathord{\left/
{\vphantom {{be^{j\phi } } {\sqrt 2 }}} \right.
\kern-\nulldelimiterspace} {\sqrt 2 }}} \right)} \right)^2 }}{{2\sigma ^2 /\Delta }}} \right\}\cdot\\
\exp \left\{ {\frac{{\left( {y_R^{{\mathop{\rm Im}\nolimits} } [2N + 1] - {\mathop{\rm Im}\nolimits} \left( {{{be^{j\phi } } \mathord{\left/
{\vphantom {{be^{j\phi } } {\sqrt 2 }}} \right.
\kern-\nulldelimiterspace} {\sqrt 2 }}} \right)} \right)^2 }}{{2\sigma ^2 /\Delta }}} \right\}.$

\subsubsection{Message Update Rules}
\begin{figure}[tt]
\centering
\includegraphics[width=0.45\textwidth]{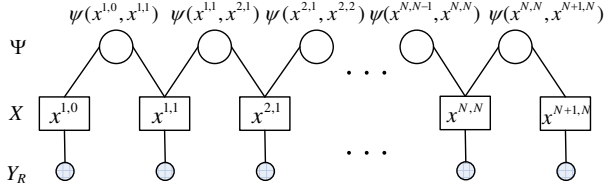}
\caption{Tanner graph of the BP-UPNC.} \label{fig:TannerBPPNC}
\end{figure}
Given the evidence node values computed by (\ref{equ:ApncBPIni1}) and (\ref{equ:ApncBPIni2}), we now derive the message update rules for BP-PNC. Since the Tanner graph (in Fig. \ref{fig:TannerBPPNC}) for BP-PNC has a tree structure, the decoding of the joint probability   can be done by passing the messages only once on each direction of an edge \cite{YedTR01}. As described below, we could consider the right-bound messages first followed by the left-bound messages.

We represent messages on the edges with respect to the compatibility nodes $\Psi$ in Fig. \ref{fig:TannerBPPNC} by $Q_k$ and $R_k$. Specifically, $Q_k$ ($R_k$) is the message connected to the right (left) of the \emph{k}-th compatibility node. $P_k  = (p_k^{1 + j,1 + j} ,p_k^{1 + j, - 1 + j} ,...,p_k^{1 - j,1 - j} )$ is a $16 \times 1$ probability vector, where each component $p_k^{a,b}$ is the joint conditional probability of $\left(x_A [\left\lceil {k/2} \right\rceil ],x_B [\left\lfloor {k/2} \right\rfloor ]\right)$ given $y_R [k]$ of a particular $(a,b)$ combination. Similarly, $Q_k  = (q_k^{1 + j,1 + j} ,q_k^{1 + j, - 1 + j} ,...,q_k^{1 - j,1 - j} )$ and $R_k  = (r_k^{1 + j,1 + j} ,r_k^{1 + j, - 1 + j} ,...,r_k^{1 - j,1 - j} )$ are also $16 \times 1$ probability vectors, where $q_k^{a,b}$ and $r_k^{a,b}$ are probabilities $P\left(x_A [\left\lceil {k/2} \right\rceil ] = {a \mathord{\left/ {\vphantom {a {\sqrt 2 }}} \right. \kern-\nulldelimiterspace} {\sqrt 2 }},x_B [\left\lfloor {k/2} \right\rfloor ] = {b \mathord{\left/ {\vphantom {b {\sqrt 2 }}} \right. \kern-\nulldelimiterspace} {\sqrt 2 }}|y_R [1],...,y_R [k - 1]\right)$ and $P\left(x_A [\left\lceil {k/2} \right\rceil ] = {a \mathord{\left/ {\vphantom {a {\sqrt 2 }}} \right. \kern-\nulldelimiterspace} {\sqrt 2 }},x_B [\left\lfloor {k/2} \right\rfloor ] = {b \mathord{\left/ {\vphantom {b {\sqrt 2 }}} \right. \kern-\nulldelimiterspace} {\sqrt 2 }}|y_R [1],...,y_R [k] \right)$, respectively, for the right-bound messages as shown in the \textbf{Step 1} (note here, for the left-bound message $q_k^{a,b}$ and $r_k^{a,b}$ are probabilities $P\left(x^{\left\lceil {k/2} \right\rceil {\rm{,}}\left\lfloor {k/2} \right\rfloor }  = {{(a,b)} \mathord{\left/ {\vphantom {{(a,b)} {\sqrt 2 }}} \right. \kern-\nulldelimiterspace} {\sqrt 2 }}|y_R [k],...,y_R [2N + 1]\right)$ and $P\left(x^{\left\lceil {k/2} \right\rceil {\rm{,}}\left\lfloor {k/2} \right\rfloor }  = {{(a,b)} \mathord{\left/ {\vphantom {{(a,b)} {\sqrt 2 }}} \right. \kern-\nulldelimiterspace} {\sqrt 2 }}|y_R [k + 1],...,y_R [2N + 1]\right)$ correspondingly).

We omit the index \emph{k} to avoid cluttering in the following discussion of message-update rules. We follow the principles and
assumptions of the BP algorithm to derive the update equations, \emph{i.e.}, the output of a node should be consistent with the inputs
while adopting a `sum of product' format of the possible input combinations \cite{YedTR01}.\vspace*{.04in}\\
% here I do not want this Step to indent
\textbf{Step 1. Update of right-bound messages}
\vspace*{.04in}
\begin{figure}
   \begin{minipage}[t]{0.45\linewidth}
     %\centering
      \includegraphics[width=1.0\textwidth]{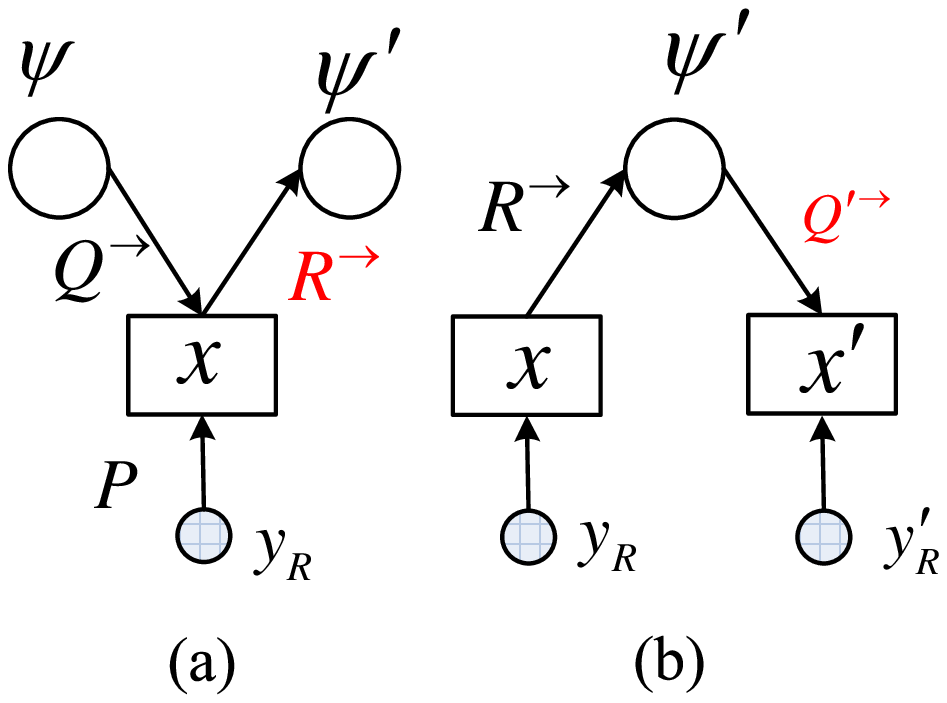}
      \caption{Messages passing from left to right.}
      \label{fig:MessageLeftToRight}
   \end{minipage}
   \hfill
   \begin{minipage}[t]{0.45\linewidth}
     %\centering
      \includegraphics[width=1.0\textwidth]{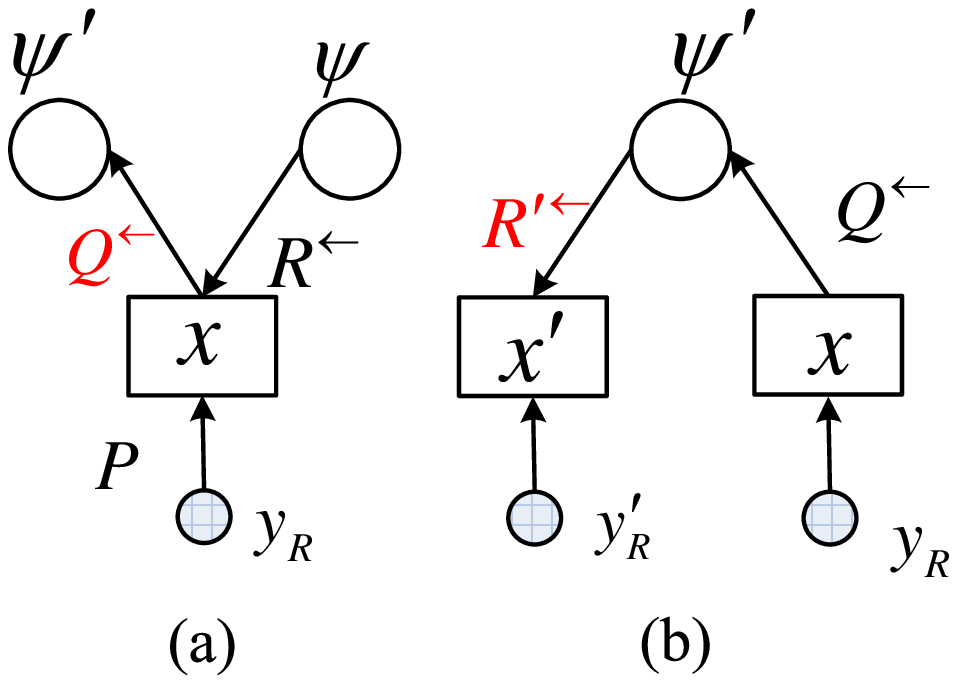}
      \caption{Messages passing from right to left.}
      \label{fig:MessageRightToLeft}
   \end{minipage}
\end{figure}

%\begin{figure}[tt]
%\centering
%\includegraphics[width=0.30\textwidth]{MessageLeftToRight.eps}
%\caption{Messages passing from left to right.} \label{fig:MessageLeftToRight}
%\end{figure}
%
%\begin{figure}[tt]
%\centering
%\includegraphics[width=0.30\textwidth]{MessageRightToLeft.eps}
%\caption{Messages passing from right to left.} \label{fig:MessageRightToLeft}
%\end{figure}

\begin{figure*}
   \begin{minipage}[t]{0.48\linewidth}
     %\centering
      \includegraphics[width=0.998\textwidth]{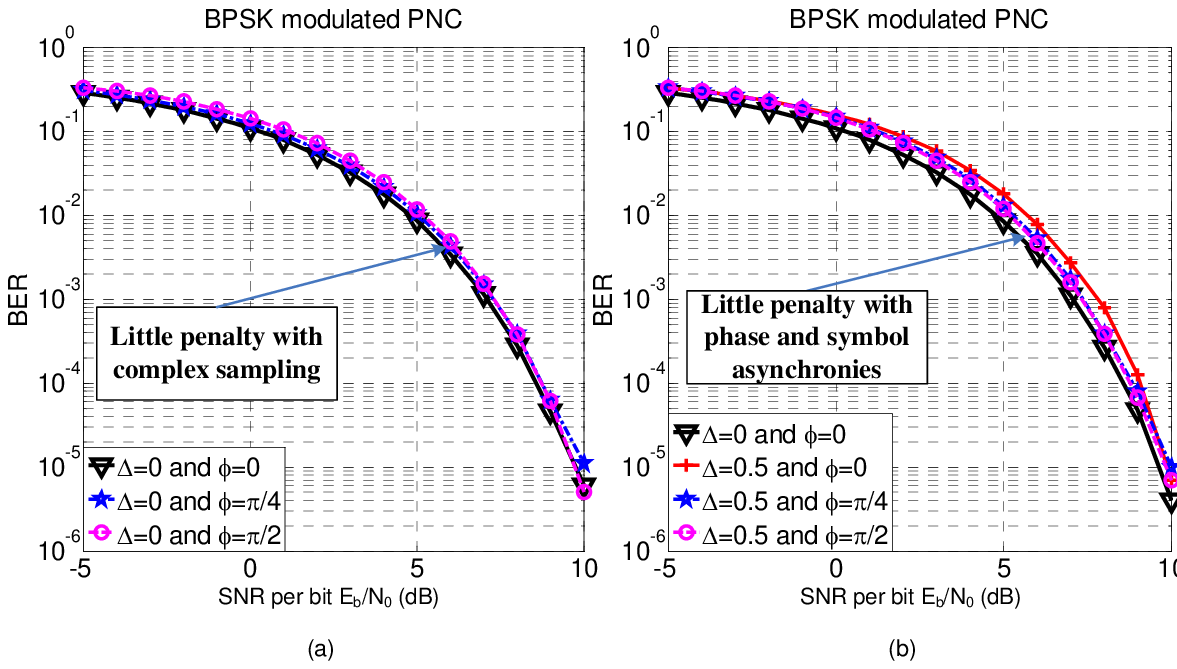}
      \caption{Simulations results of BP-PNC with BPSK modulation.}
      \label{fig:BPPNCsimulationBPSK}
   \end{minipage}
   \hfill
   \begin{minipage}[t]{0.48\linewidth}
     \centering
      \includegraphics[width=1.05\textwidth]{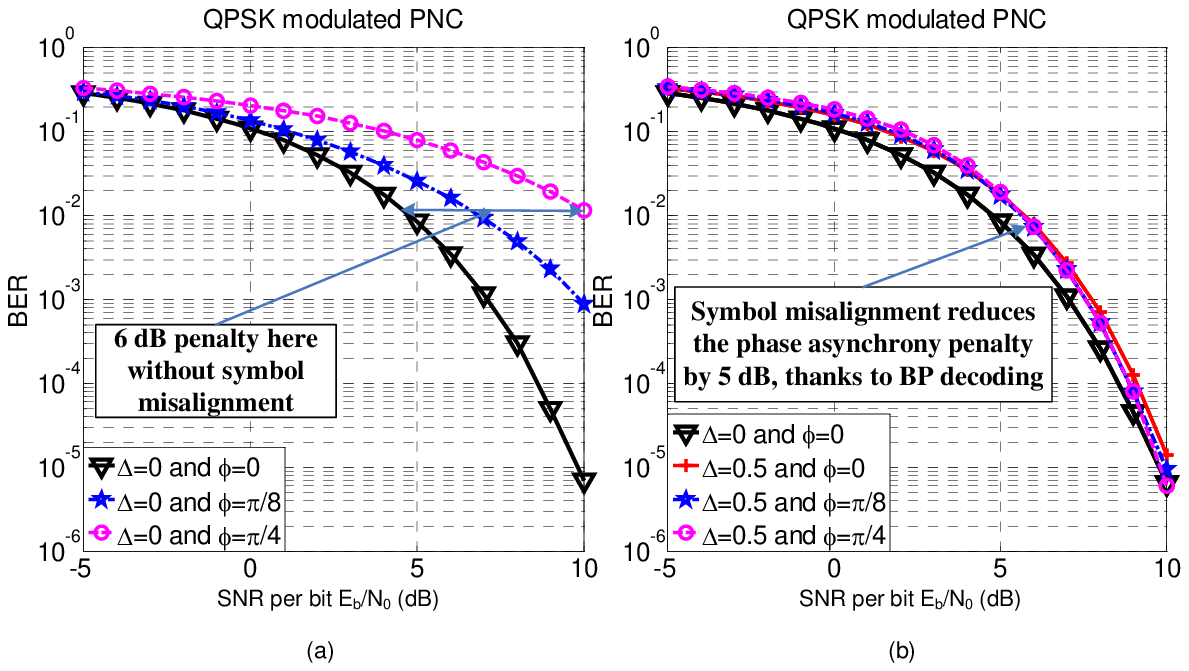}
      \caption{Simulations results of BP-PNC with QPSK modulation.}
      \label{fig:BPPNCsimulationQPSK}
   \end{minipage}
\end{figure*}

With reference to Fig. \ref{fig:MessageLeftToRight}(a), we derive the update equations for the right-bound message $R^ \to   = (r^{1 + j,1 + j} ,r^{1 + j, - 1 + j} ,...,r^{1 - j,1 - j} )$ based on the right-bound message  and the fixed message $Q^ \to   = (q^{1 + j,1 + j} ,q^{1 + j, - 1 + j} ,...,q^{1 - j,1 - j} )$ from the evidence node $P = (p^{1 + j,1 + j} ,p^{1 + j, - 1 + j} ,...,p^{1 - j,1 - j} )$. Based on the sum-product principle of the BP algorithm, we have
\begin{align}
r^{a,b}  = p^{a,b} q^{a,b}. \label{equ:ApncBPUpdate1}
\end{align}
For the input message into the leftmost compatibility node, however, we should modify (\ref{equ:ApncBPUpdate1}) with $r^{a,b}  = p^{a,b}$.

With reference to Fig. \ref{fig:MessageLeftToRight}(b) we derive the update equations for the message ${Q}'^\to   = (q'^{1 + j,1 + j} ,q'^{1 + j, - 1 + j} ,...,q'^{1 - j,1 - j} )$ based on the input message $R^ \to = (r_{}^{1 + j,1 + j} ,r_{}^{1 + j, - 1 + j} ,...,r_{}^{1 - j,1 - j} )$ of the compatibility node. Note that for $x = x^{n,n}$ and $x' = x^{n + 1,n}$, the common symbol overlapping in two adjacent samples is $x_B [n]$, thus we have
\begin{align}
q'^{1 + j,b}  = q'^{ - 1 + j,b}  = q'^{ - 1 - j,b}  = q'^{1 - j,b} = \beta \sum\limits_{a \in \chi } {r^{a,b} }, \label{equ:ApncBPUpdate2}
\end{align}
where $b \in \chi $ and $\beta$ is a normalization factor to ensure that the sum of probabilities $\sum\limits_{a,b \in \chi} {q'^{a,b} }  = 1$.
If $x = x^{n + 1,n}$ and $x' = x^{n + 1,n + 1}$, then the common symbol is $x_A [n + 1]$. We have
\begin{align}
q'^{a,1 + j}  = q'^{a, - 1 + j}  = q'^{a, - 1 - j}  = q'^{a,1 - j} = \beta \sum\limits_{b \in \chi } {r^{a,b} }, \label{equ:ApncBPUpdate3}
\end{align}
where $a \in \chi$ and $\beta$ is a normalization factor to ensure that the sum of probabilities $\sum\limits_{a,b \in \chi} {q'^{a,b} }  = 1$.

We then update the next evidence node message $R'^\to$  based on the newly updated $Q'^\to$ in Fig. \ref{fig:MessageLeftToRight}(b) and $y'_R $ in Fig. \ref{fig:MessageLeftToRight}(a), and so on and so forth, until we reach rightmost node.\vspace*{.04in}\\
\textbf{Step 2. Update of left-bound messages}
\vspace*{.04in}

With reference to Fig. \ref{fig:MessageRightToLeft}(a) and (b), we use a similar procedure as in \textbf{Step 1} to update the left-bound messages.

%After steps 1 and 2 above, the values of the messages converge, thanks to the tree structure of the Tanner graph. The $n$-th source message $M_n$ (a $16 \times 1$ probability vector) can be calculated as $M_{n} = P_{n} \cdot Q_{n}^{\rightarrow} \cdot R_{n}^{\leftarrow}$ , where $P$, $Q^\to$ and $R^ \leftarrow$ are the three messages flowing into the even node $x^{n,n}$ in $X$. We then map $M_{n}$ to a $4 \times 1$ vector $\tilde M_{n}$ based on synchronous QPSK PNC mapping\footnote{Since we have already incorporate the phase and symbol asynchronies into source nodes by the BP decoding, we can use QPSK PNC mapping here.} (\emph{e.g.} the first element $\tilde m_{n}^{1+j} = P\left( x_R[n] = 1+j | Y_R \right) = \sum\limits_{a \in \chi } {\left( {p_k^a q_k^a r_k^a } \right)}$). For ML detection, we declare that $x_{R}[n] = \left( {x_A [n] \oplus x_B [n]} \right) = \arg \mathop {\max }\limits_{(b):b \in \chi } \left( m_n^b \right)$, where $p_{2n}^{a,b}$, $q_{2n}^{a,b}$ and $r_{2n}^{a,b}$ are obtained from the three messages $P$, $Q^\to$ and $R^ \leftarrow$.
%$\left( {\sum\limits_{a - b = 0} {\left( {p_{2n}^{a,b}  \cdot q_{2n}^{a,b}  \cdot r_{2n}^{a,b} } \right)} ,\sum\limits_{a - b =  \pm 2j} {\left( {p_{2n}^{a,b}  \cdot q_{2n}^{a,b}  \cdot r_{2n}^{a,b} } \right)} ,\sum\limits_{a - b =  \pm 2} {\left( {p_{2n}^{a,b}  \cdot q_{2n}^{a,b}  \cdot r_{2n}^{a,b} } \right)} ,\sum\limits_{a - b =  \pm 2 \pm 2j} {\left( {p_{2n}^{a,b}  \cdot q_{2n}^{a,b}  \cdot r_{2n}^{a,b} } \right)} } \right)$
\vspace*{.06in}
After steps 1 and 2 above, the values of the messages converge, thanks to the tree structure of the Tanner graph. We then compute the 4-tuple.
\begin{align}
\left(\sum\limits_{(a,b):a \oplus b = 1 + j} {p_{2n}^{a,b} q_{2n}^{a,b} r_{2n}^{a,b} } , \sum\limits_{(a,b):a \oplus b =  - 1 + j} {p_{2n}^{a,b} q_{2n}^{a,b} r_{2n}^{a,b} }, \right.\nonumber\\
\left.\sum\limits_{(a,b):a \oplus b =  - 1 - j} {p_{2n}^{a,b} q_{2n}^{a,b} r_{2n}^{a,b} } , \sum\limits_{(a,b):a \oplus b = 1 - j} {p_{2n}^{a,b} q_{2n}^{a,b} r_{2n}^{a,b} } \right), \nonumber
\end{align}
where $a,b \in \chi$. The maximum-likelihood ${x_A [n] \oplus x_B [n]}$ is given by the $a \oplus b$ that yields the largest element among the four elements. Note that we only use $P\left( {x_A [n],x_B [n]} | Y_R \right)$, and not $P\left( {x_A [n + 1],x_B [n]} | Y_R \right)$, to obtain $P\left( {x_A [n] \oplus x_B [n]} | Y_R \right)$, because the information required to get $P\left( {x_A [n] \oplus x_B [n]} | Y_R \right)$ is fully captured in $P\left( {x_A [n],x_B [n]} | Y_R \right)$.

\section{Numerical Results}
This section presents simulation results for BP-PNC. We compare the performance of asynchronous PNC with that of the perfectly synchronized case \cite{PNC06}, focusing on BPSK and QPSK modulations. For comparison purposes, the transmit amplitude per symbol in the QPSK case is made to be $\sqrt {2}$ of that in the BPSK case so that they have the same energy per bit. In particular, $E_b$ is the energy per bit in the Fig. \ref{fig:BPPNCsimulationQPSK}. The ideal synchronous QPSK curve therefore overlaps with the synchronous BPSK curve (because when symbols and phase are synchronized in QPSK, the system can be considered as two parallel BPSK systems).
\vspace*{-0.11in}
\subsection{Summary of results}
Our simulations yield the following findings:
\begin{itemize}
\item For BPSK, the 3 dB BER performance penalty due to phase or symbol asynchrony using the decoding methods in \cite{PNC06, SyncPNC06} is reduced to less than 0.5 dB with our method here.
\item For QPSK, the BER performance penalty due to phase asynchrony can be as high as 6-7 dB when the symbols are aligned, as can be seen from Fig. \ref{fig:BPPNCsimulationQPSK}(a). However, with half symbol misalignment, our BP decoding algorithm can reduce the penalty to less than 1 dB, thanks to a ``diversity and certainty propagation'' effect (to be explained).
\item Symbol misalignment makes the system more robust against phase asynchrony under our ML decoding method.
\end{itemize}
\vspace*{-0.11in}
\subsection{Detailed description}\label{CertProp}
Fig. \ref{fig:BPPNCsimulationBPSK} and Fig. \ref{fig:BPPNCsimulationQPSK} show the simulation results for BPSK and QPSK modulations, respectively. The \emph{x}-axis is the average per-bit SNR of both end nodes, and the \emph{y}-axis is the average $BER = {{\left( {BER_A  + BER_B } \right)} \mathord{\left/ {\vphantom {{\left( {BER_A  + BER_B } \right)} 2}} \right. \kern-\nulldelimiterspace} 2}$.

For each data point, we simulate 10,000 packets of 2,048 bits. We use the synchronous PNC as a benchmark to evaluate BP-PNC. As can be seen from Fig. \ref{fig:BPPNCsimulationBPSK}(a) and (b), the BER performance penalties due to phase and symbol asynchronies are less than 0.5 dB for all SNR regimes for BPSK. That is, for BPSK, BP-PNC reduces the BER performance penalty from 3dB in \cite{PNC06} to only 0.5 dB. Note that BP-PNC uses complex sampling that samples the 2-D information in the In-phase and Quadrature-phase components of the signal, whereas the scheme in \cite{PNC06} only samples in one dimension. For optimality, complex sampling is required in BPSK PNC because of the phase offset.

For QPSK with symbol synchrony but phase asynchrony, as can be seen from Fig. \ref{fig:BPPNCsimulationQPSK}(a), the BER performance penalty can be as large as 6 to 7 dB. However, with half symbol misalignment, as can be seen from Fig. \ref{fig:BPPNCsimulationQPSK}(b), BP-PNC reduces the penalty to within 1 dB (compared with the benchmark case where symbol and phase are perfectly synchronized). In other words, symbol asynchrony can ameliorate the penalty due to phase asynchrony. This can be explained by the ``diversity and certainty propagation'' effects elaborated in the next subsection. In addition, we note from Fig. \ref{fig:BPPNCsimulationQPSK}(b) that when there is a half symbol offset, the phase offset effect becomes much less significant. Specifically, the spread of SNRs for a fixed BER under different phase offsets is less than 0.5 dB.
\vspace*{-0.11in}
\subsection{Diversity and Certainty Propagation}
\begin{figure}[tt]
\centering
\includegraphics[width=0.40\textwidth]{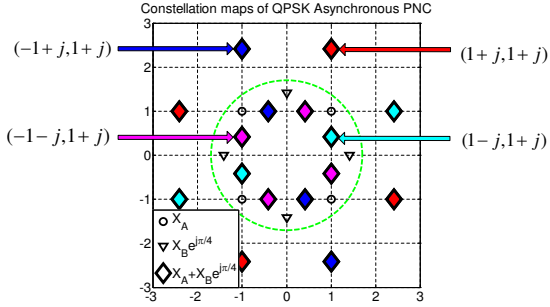}
\caption{Constellation map for phase asynchronous QPSK PNC (with symbol synchronization). There are four \emph{XOR}ed PNC symbols for QPSK. Constellation points of the same color in the figure should be mapped to a same PNC symbol. And the amplitude of each symbol is $\sqrt 2$.} \label{fig:ConstMapQPSKAPNC}
\end{figure}

\begin{figure}[tt]
\centering
\includegraphics[width=0.35\textwidth]{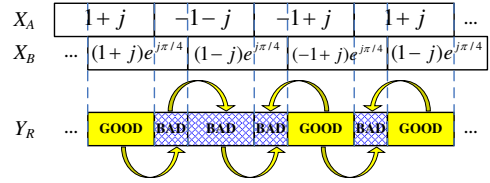}
\caption{An illustration of certainty propagation. The yellow symbols denote the good constellation points and the white symbols with grids denote the bad constellation points of Fig. \ref{fig:ConstMapQPSKAPNC}, respectively.} \label{fig:BPPNCCertProp}
\end{figure}

In QPSK, each symbol has four possible values. Thus, the joint symbol from both sources has 16 possible values. The constellation map of the joint symbol varies according to the phase offset. Fig. \ref{fig:BPPNCCertProp} shows the constellation map of a joint symbol when the phase offset is $\pi/4$, where the 16 diamonds corresponding to the 16 possibilities. For example, a point with value $1 + (1 - \sqrt 2 )j$ corresponds to the joint symbol $x^{n,n}  = \left( {1 + j, - 1 - j} \right)$ in Fig. \ref{fig:ConstMapQPSKAPNC} due to the  phase shift (\emph{i.e.}, $1 + (1 - \sqrt 2 )j = (1 + j) + ( - 1 - j)e^{j{\pi  \mathord{\left/ {\vphantom {\pi  4}} \right. \kern-\nulldelimiterspace} 4}}$). In PNC, the 16 possibilities need to be mapped to four \emph{XOR} possibilities for the PNC symbol. In Fig. \ref{fig:ConstMapQPSKAPNC}, the diamonds are grouped into groups of four different colors. The diamonds of the same color are to be mapped to the same \emph{XOR} PNC symbol according to $x_A  \oplus x_B$. In this mapping process, some of the constellation points are more prone to errors than other constellation points, and the BER is dominated by these bad constellation points.

With reference to Fig. \ref{fig:ConstMapQPSKAPNC}, the eight diamonds within the green circle are ``bad constellation points''. Adjacent points among the 8 points are mapped to different \emph{XOR} values, but the distances among two adjacent points are short. By contrast, the eight points outside the green circle are ``good constellation points'' because the distances among adjacent points are large. When symbols are synchronized (\emph{i.e.}, $\Delta  = 0$), there are altogether $N$ joint symbols. On average, half of them will be bad constellation points with high BER.

Now, consider what if there is symbol asynchrony, say $\Delta  = 0.5$. We have $2N+1$ joint symbols, out of which about $N$ will be good constellation points. A symbol from a source is combined with two symbols from the other sources to form two joint symbols received at the relay. Both the joint symbols have to be bad for poor performance. Thus, the diversity itself may give some improvement. In addition, there is a certainty propagation effect, as explained below.

Consider a good constellation point, say $x^{n,n}  = \left( {x_A [n],x_B [n]} \right)$. For this point, we may be able to decode not just the \emph{XOR}, but the individual values of $x_A [n]$ and $x_B [n]$ with high certainty. Now, suppose that the next joint symbol $x^{n{\rm{ + 1}},n}  = (x_A [n{\rm{ + 1}}],x_B [n])$ is a bad constellation point. But since we have good certainty about $x_B [n]$ from the previous good constellation point, the uncertainly in $x^{n + 1,n}$ can be reduced by the certainty propagated from $x^{n,n}$ with the BP algorithm. In other words, once $x_B [n]$ is known, $x^{n + 1,n}$ is not a bad constellation point anymore. Certainty can propagate along successive symbols from left to right, as well as from right to left, as shown in Fig. \ref{fig:BPPNCCertProp}. Certainty propagation significantly reduces BER.

%To summarize, with the help of BP-PNC, the SNR penalty is reduced to within 1 dB for cases with both phase and symbol asynchrony.

\section{Conclusion and Future Work}
This paper has proposed and investigated an optimal maximum likelihood (ML) decoding algorithm based on belief propagation, called BP-PNC, for asynchronous PNC. This method significantly improves the BER performance of the prior methods when there are symbol and phase asynchronies in the PNC system. In addition, we find that with symbol misalignment, BP-PNC desensitizes the BER performance to carrier-phase offset. In practice, it is relatively easier to control symbol timing than carrier-phase offset because of the relative time scales involved. Our results suggest that if we could control the symbol offset in PNC (e.g., using the method in \cite{RahulSourceSync10}), it would actually be advantageous to deliberately introduce symbol misalignment so that the system is robust against uncontrollable phase offsets.

Prior to this work, it has often been thought that strict synchronization is needed for PNC. Our work, however, suggests that the penalty due to asynchrony can be nullified to a large extent with the BP-PNC decoding method. Our work is based on BPSK and QPSK modulations. We believe the same conclusion can be drawn for other modulations, particularly those with dense constellations. In addition, channel coding (with or without symbol offset in the system) also has the diversity and certainly propagation effects of symbol offset expounded in Section \ref{CertProp}. Thus, we believe that the incorporation of channel coding will also lessen the negative effect of asynchrony in PNC systems. These conjectures remain to be investigated in future work.

%Our simulation results reveal that BP-PNC reduces the SNR penalty due to phase asynchrony. Specifically, while exploiting the symbol asynchrony, BP-PNC can better perform the symbol synchronous UPNC by around 2.5 dB for BPSK and 6 dB for QPSK, respectively. So  It is known that, to control the phase shift is relatively harder than the symbol timing. Imagine that we can control the symbol misalignment [RahulSourceSync10] \cite{RahulSourceSync10}, and then we should intentionally have half-symbol misalignment so that we have a robust system that is immunized to the phase asynchrony.

%\begin{figure}[tt]
%\centering
%\includegraphics[width=0.465\textwidth]{PERcomparison1.eps}
%\caption{PER Comparison of C-CRESM and Turbo-SIC for different
%$\Delta$.} \label{fig:PERcomparison1}
%\end{figure}

% conference papers do not normally have an appendix

%% use section* for acknowledgement
%\section*{Acknowledgment}
%
%
%The authors would like to thank...

% that's all folks
\end{document}